\documentclass[12pt]{article}

\usepackage{geometry} 
\usepackage{graphicx} 
\usepackage[colorlinks,linkcolor=blue,anchorcolor=blue,citecolor=blue]{hyperref}
\usepackage{float} 
\usepackage{wrapfig} 
\usepackage{lipsum} 
\usepackage{exscale}
\usepackage{relsize}
\usepackage{indentfirst}
\usepackage{amssymb}
\usepackage{tikz}
\usetikzlibrary{positioning, calc, math, matrix}
\usepackage[linesnumbered,lined,boxed,commentsnumbered]{algorithm2e}
\usepackage{lscape}
\usepackage[inline]{enumitem}

\usepackage{makecell}
\usepackage{amsthm}
\usepackage{amsmath,bm}
\usepackage{abstract}
\usepackage{setspace}
\usepackage[figuresright]{rotating}
\usepackage{threeparttable}
\usepackage[misc]{ifsym}
\usepackage{anyfontsize}

\tikzset{baseline={([yshift=-0.5ex]current bounding box.center)},
         every matrix/.style={matrix of math nodes, column sep=3em, row sep=3em}}

\DeclareMathOperator{\Tr}{Tr}

\renewcommand{\arraystretch}{1.8}

\numberwithin{equation}{section}

\newtheorem{thm}{Theorem}[section]
\newtheorem{defn}[thm]{Definition}
\newtheorem{lem}[thm]{Lemma}
\newtheorem{coro}[thm]{Corollary}

\newtheorem{remark}[thm]{Remark}
\newtheorem{exam}[thm]{Example}

\begin{document}
\title{New constructions of DNA codes under multiple constraints and parallel searching algorithms\let\thefootnote\relax\footnotetext{
 E-Mail addresses: wangguodong@mails.ccnu.edu.cn (G. Wang),\\ hwliu@ccnu.edu.cn (H. Liu), xueyanchen@mails.ccnu.edu.cn(X. Chen)
}
}

\author{~Guodong Wang,~Hongwei Liu,~Xueyan Chen}

\date{\small School of Mathematics and Statistics, Central China Normal University, Wuhan, 430079, China \\}
\maketitle

\begin{abstract}

DNA codes have garnered significant interest due to their utilization in digital media storage, cryptography, and DNA computing.
In this paper, we first extend the results of constructing reversible group codes \cite{Cengellenmis} and reversible composite group codes \cite{Korban5} to general even-order finite groups.
By using these results, we give parallel searching algorithms to find some new DNA codes with better parameters.
Secondly, by mapping codes over $\mathbb{F}_4$ to DNA codes,
we establish a relationship between the $GC$-weight enumerator of DNA codes and the Hamming weight enumerator of their trace codes,
which greatly improves the computational efficiency of searching for DNA codes.
Based on this relationship, we propose an efficient algorithm for generating DNA codes with $50\%$ $GC$-content.
Furthermore, we find that there is no direct connection between the $GC$-weight enumerator of a DNA code and the $GC$-weight enumerator of its dual code.
Finally, we present algorithms for determining whether a DNA code is free from secondary structures or conflict-free,
and some new DNA codes with better parameters under multiple constraints are obtained, which are listed in Tables 1 and 4.

\noindent \textbf{Keywords:}
DNA codes. reversible group codes. quasi-cyclic codes

\end{abstract}

\newpage

\section[Introduction]{Introduction}
DNA molecules are composed of two complementary strands,
each containing a sequence of four distinct nucleotide bases: adenine $(A)$, cytosine $(C)$, guanine $(G)$, and thymine $(T)$.
DNA molecules have a double-helix structure,
in which adenine pairs with thymine ($A-T$),
and cytosine pairs with guanine ($C-G$).
We use the notation $A^c=T,T^c=A,C^c=G$, and $G^c=C$
to represent the Watson-Crick complement of each nucleotide base.
This complementary pairing is the basis for the stable structure and functional realization of DNA molecules.
Recently, DNA codes have garnered significant interest due to their utilization in digital media storage, cryptography and DNA computing.
For instance, see \cite{Adleman, Boneh, Blawat, Church}.

It is well known that a good DNA code satisfies the following constraints:
(1) the Hamming distance constraint, (2) the reverse constraint,
(3) the reverse-complement constraint, and (4) the fixed $GC$-content constraint,
\cite{Aboluion1}, \cite{Abualrub}, \cite{Benerjee}, \cite{Gaborit}, \cite{Zhang}.

In addition to the above constraints,
there are two other crucial constraints that have been studied in the literature;
being free from secondary structures and being conflict-free,
for example, see \cite{Bornholt}, \cite{Benerjee}, \cite{Kari}, and \cite{Kova}.
Secondary structure formation is a major bottleneck in DNA storage systems and  DNA computing.
In \cite{Benerjee}, \cite{Kari}, \cite{Milenkovic}, and \cite{Nelms}, the authors studied conflict-free DNA strings,
one should avoid having two sub-strings
where the reverse complement of one sub-string is the other one
because this structure results in
forming an anti-parallel double stranded hairpin like structure
by folding back upon itself.

Studies on cyclic and extended cyclic constructions \cite{Aboluion1},
\cite{Abualrub} as well as on linear constructions of DNA codes \cite{Gaborit} have been conducted.
In \cite{Cengellenmis}, \cite{Dougherty6} and \cite{Korban5},
the authors considered linear codes derived from group ring elements for constructing reversible DNA codes that satisfy certain constraints.
Consequently, numerous new lower bounds on the sizes of specific DNA codes have been discovered.
Recently, in \cite{Dougherty8}, some DNA codes have been constructed using special groups.
In this paper, we further explore the construction of DNA codes utilizing group codes and composite group codes.
For the convenience of discussion,
we only consider the case where the length of DNA codes is even.

We first extend the results of constructing reversible group codes \cite{Cengellenmis} and reversible composite group codes \cite{Korban5} to general even-order finite groups.
This significantly increases the number of groups that can be utilized in constructing DNA codes. Based on this result,
we further propose an efficient parallel searching algorithm aimed at accelerating the search process for DNA codes.
By incorporating existing optimal results as input parameter,
we significantly enhance the efficiency of finding DNA codes with better parameters.

The $50\%$ $GC$-content is a crucial metric for DNA codes.
When considering DNA codes as codes over $\mathbb{F}_4$,
we establish a direct relationship between the $GC$-weight enumerator of a DNA code and the Hamming weight enumerator of its trace code.
This relationship enables us to more rapidly construct DNA codes that satisfy the $50\%$ $GC$-content constraint.
Moreover, by combining this relationship with the properties of Krawtchouk polynomials,
we can facilitate the faster screening of DNA codes that meet the $50\%$ $GC$-content constraint.
The specific improvements in computational efficiency can be observed in Tables 2 and 3.

The paper is organized as follows.
In Section 2, we elaborate on the basic concepts of DNA codes,
describe various constraints of DNA codes,
and introduce other concepts related to DNA codes.
In Section 3, we extend the results of constructing DNA codes using group codes and composite group codes to general finite groups of even order,
which significantly expands the range of groups we can choose from when constructing DNA codes.
In Section 4, we conduct an in-depth study on the $GC$-weight enumerator of DNA codes and
establish a relationship between the $GC$-weight enumerator of a DNA code and the Hamming weight enumerator of its trace code,
which facilitates finding DNA codes with better parameters using computer programs.
Furthermore, we provide examples to illustrate that,
in general, there is no direct connection between the $GC$-weight enumerator of a DNA code and the $GC$-weight enumerator of its dual code.
In Section 5, we present an efficient algorithm for finding DNA codes with better parameters,
and using this algorithm, we discover some new DNA codes with better parameters.
Additionally, we provide several algorithms tailored to various constraints on DNA codes,
enabling us to search for better DNA codes under various constraints.
Finally, we briefly summarize our work and discuss future research directions.

All computations in this paper are performed using MAGMA \cite{Bosma}, version 2.25-3 for Windows.
Most tasks are executed on a computer with an Intel Xeon CPU E3-1231 v3 @ 3.40GHz processor.
All the data we obtained and most of the algorithms presented in this paper can be found in \cite{OurData}.

\section{Preliminaries}
A DNA code $D$ of length $n$ is defined as a set of codewords $\mathbf{x}= (x_1,x_2,\ldots,x_n)$, where $x_i\in S_{D_4}=\{A,T,C,G\}$,
such that $D$ satisfies some or all of the following constraints:

(i) The Hamming distance constraint (HD):
$$d(\mathbf{x},\mathbf{y})\geq d,\quad\forall\:\mathbf{x},\mathbf{y}\in D,$$
for some Hamming distance $d.$

(ii) The reverse constraint (R):
$$d(\mathbf{x}^r,\mathbf{y})\ge d,\quad\forall\:\mathbf{x},\mathbf{y}\in D,\:\text{including}\:\mathbf{x}=\mathbf{y}$$
for some Hamming distance $d.$

(iii) The reverse-complement constraint (RC):
$$d(\mathbf{x}^{rc},\mathbf{y})\geq d,\:\forall\:\mathbf{x},\mathbf{y}\in D,\:\text{including}\:\mathbf{x}=\mathbf{y}$$
for some Hamming distance $d.$

(iv) The fixed $GC$-content constraint (GC): The set of
codewords with length $n$, distance $d$ and $GC$-weight $w_{gc}$,
where $w_{gc}$ is the total number of $G$'s and $C$'s present in the DNA strand:
$$w_{gc}(\mathbf{x})={\Big|}\Big \{x_i:\mathbf{x}=(x_i),x_i\in\{C,G\} \Big\}{\Big |},$$
where
\begin{align*}
    \mathbf{x}^r &= (x_n,x_{n-1},\ldots,x_2,x_1) & \text{(Reverse of $\mathbf{x}$)} \\
    \mathbf{x}^c &= (x_1^c,x_2^c,\ldots,x_n^c) & \text{(Complement of $\mathbf{x}$)} \\
    \mathbf{x}^{rc} &= (x_n^c,x_{n-1}^c,\ldots,x_2^c,x_1^c) & \text{(Reverse complement of $\mathbf{x}$)}
\end{align*}

In this paper, we simply fix the $GC$-content to be half the length of the DNA
code $D$.

A DNA string is deemed free from secondary structures (stem length $> 2$) if it lacks any reverse-complement substrings longer than $2$,
equivalent to being clear of structures with stem length $3$.
Hence, we focus on this length $3$ property.

For integers $n$ and $l \leq n/2$, a DNA string of length $n$ is called $l$ conflict-free if it avoids consecutive repetitions of identical substrings up to length $l$ \cite{Benerjee}.
For example, $ACTGACTGTGAC$ is $3$ conflict-free but not $4$ conflict-free,
due to the repeated $4$-length substring $ACTG$.
It's also known \cite{Song} that an $l$ conflict-free string is inherently $l-1$ conflict-free for $l \geq 2$.
A DNA code of length $n$ is $l$ conflict-free if all its codewords meet this criterion.
Generally, insertion/deletion errors are prevalent in strings with consecutive nucleotide/block repetitions,
causing misalignment during sequencing.
Thus, DNA codes excluding such repetitions are favored in applications.

A DNA code can be identified with a code over the finite field $\mathbb{F}_4=\{0, 1, w, w^2\}$,
where $w^2=w+1$. This identification is achieved by employing the standard bijective correspondence between $\mathbb{F}_4$ and the DNA alphabet $S_{D_4}=\{A,T,C,G\}$,
which is given by
\[
    \eta:\mathbb{F}_4 \rightarrow S_{D_4},
\]
with $\eta(0)=A, \eta(1)=T,\eta(w)=C$, and $\eta(w^2)=G$.
Thus, in this paper we do not distinguish between codes over $\mathbb{F}_4$ and DNA codes.
We denote the complete weight enumerator of a code $C$ over the binary field $\mathbb{F}_2$ by
\[
    CWE_C (a, b) = \sum_{\mathbf{c} \in C} a^{n_0(\mathbf{c})}b^{n_1(\mathbf{c})}.
\]
We denote the complete weight enumerator of a code $C$
over $\mathbb{F}_4$ by
\[
    CWE_C (a, b, c, d) = \sum_{\mathbf{c} \in C} a^{n_0(\mathbf{c})}b^{n_1(\mathbf{c})}c^{n_w(\mathbf{c})}d^{n_{w^2}(\mathbf{c})},
\]
where $n_s(\mathbf{c})$ denotes the number of occurrences of $s$ in a codeword $\mathbf{c}$.
We identify the complete weight enumerator of a
DNA code $D$ with that of a code $C$ over $\mathbb{F}_4$ where $D = \eta(C)$.
Therefore, if we let
\[
    GCW_C(a, b)=CWE_C (a, a, b, b),
\]
then $GCW_C (a, b)$ is the $GC$-weight enumerator of a code
$C$, where the coefficient of $b^i$ is the same as the number of
codewords with $GC$-weight $i$.

We briefly introduce the structure of quasi-cyclic codes,
for a detailed description of quasi-cyclic codes, please refer to \cite{LingS1,LingS2}.
A linear code $C$ with length $n=ml$ is said to be a quasi-cyclic code of index $l$ and cyclic length $m$ if and
only if for any codeword
$$
(c_{0,0},\ldots,c_{0,m-2},c_{0,m-1},\ldots,c_{l-1,0},\ldots,c_{l-1,m-2},c_{l-1,m-1}) \in C,
$$ we have the shifted codeword
$$(c_{0,m-1},c_{0,0},\ldots,c_{0,m-2},\ldots,c_{l-1,m-1},c_{l-1,0},\ldots,c_{l-1,m-2}) \in C.$$
Thus, a cyclic code can be seen as a quasi-cyclic code of index $1$.

Let $R$ be a finite commutative Frobenius ring and $C$ be a quasi-cyclic code over $R$ of length $lm$ with cyclic length $m$.
Let $M =R[x]/(x^m-1)$ be the quotient ring and
$$
\mathbf{c}=(c_{0,0},\ldots,c_{0,m-1},c_{10},\ldots,c_{1,m-1},\ldots,
c_{l-1,0},\ldots,c_{l-1,m-1})
$$
denote a codeword in $C$. Define a map $\phi:R^{lm} \to M^l$ by
$$
\phi(\mathbf{c})=(c_0(x),c_1(x),\ldots,c_{l-1}(x))\in M^l, $$
where
$c_i(x)=\sum_{j=0}^{m-1}c_{ij}x^j\in M$.
Let $\phi(C)$ denote the image of $C$ under $\phi$.
It is well known that $C$ is quasi-cyclic of index $l$ and cyclic length $m$ if and only if $\phi(C)$ is an $M$-submodule of $M^l$,
where an $M$-submodule refers to a subset of $M^l$ that is closed under addition and multiplication by elements of $M$.

\section{Group codes and composite matrix group codes}

In \cite{Cengellenmis}, \cite{Korban5} and \cite{Dougherty6},
the authors utilized group codes and composite group codes to create reversible group codes for specific groups and for a
particular ordering of the group elements.
However, these two methods imposed stringent requirements on the groups.
Through further investigation, we have discovered novel approaches that can relax these conditions,
particularly by employing even-order finite groups to construct reversible group codes or reversible composite group codes.

In this section, we assume $R$ to be a finite commutative Frobenius ring.
Now, we shall give the standard definition of group rings.
Let $G$ be a finite group of order $n$.
Then any element in $RG$ is of the form $v=\sum_{i=1}^{n}\alpha_{i}g_{i}, \alpha_{i}\in R, g_{i}\in G$.
The addition in $RG$ is defined by coordinate addition, namely
\[
    \sum_{i=1}^n\alpha_ig_i + \sum_{i=1}^n\beta_ig_i = \sum_{i=1}^n(\alpha_i + \beta_i)g_i.
\]
The product of two elements in $RG$ is given by
\[
    \left(\sum_{i=1}^n\alpha_ig_i \right) \left( \sum_{j=1}^n\beta_jg_j\right) = \sum_{i,j}^n\alpha_i \beta_jg_ig_j.
\]
This gives that the coefficient of $g_k$ in the product is $\sum_{g_ig_j=g_k}\alpha_i \beta_j$.
$RG$ is called a group ring under the above two operations.

We have an $R$-linear isomorphism,
\begin{align*}
    \Omega:RG & \longrightarrow R^n,\\
    v= \alpha_{g_1} g_1 + \alpha_{g_2} g_2 + \cdots + \alpha_{g_n} g_n & \longmapsto (\alpha_{g_1},\alpha_{g_2},...,\alpha_{g_n}),
\end{align*}
for any $v=\sum_{i=1}^n \alpha_{g_i} g_i \in RG$. And we have $\Omega(gv) = (\alpha_{g^{-1}g_1},\alpha_{g^{-1}g_2},...,\alpha_{g^{-1}g_n})$,
where $g \in G$.
More precisely, this $R$-linear isomorphism is under a specific order of elements of the group $G$,
denoted by $\{g_1,g_2,...,g_n\}$.

The following matrix construction was given by Hurley in \cite{Hurley}.

\begin{defn}
Let  $G = \{g_1, g_2,..., g_n\}$ be a group of order $n$ and
$v=\sum_{i=1}^n \alpha_{g_i} g_i \in RG$.
Then
\[
    C(v)=\langle \sigma(v) \rangle
\]
is called a group code (or, for simplicity, a $G$-code) generated by $v$, where $\langle \sigma(v) \rangle$
denotes the row space of the matrix $\sigma(v)$ over $R$, given by
\setlength{\arraycolsep}{1pt}
\renewcommand\arraystretch{0.8}
\begin{align}\label{gromatr3}
    \sigma(v)=\begin{pmatrix}
        \alpha_{g_1^{-1}g_1}&\alpha_{g_1^{-1}g_2}&\ldots&\alpha_{g_1^{-1}g_n} \\
        \alpha_{g_2^{-1}g_1}&\alpha_{g_2^{-1}g_2}&\ldots&\alpha_{g_2^{-1}g_n} \\
        \vdots&\vdots& &\vdots \\
        \alpha_{g_n^{-1}g_1}&\alpha_{g_n^{-1}g_2}&\ldots&\alpha_{g_n^{-1}g_n}
    \end{pmatrix}\in M_n(R).
\end{align}
Here, $M_n(R)$ denotes the set of $n \times n$ matrices over $R$.
\end{defn}

We have $\Omega(\langle v \rangle) = \langle \sigma(v) \rangle$,
where $\langle v \rangle$ is a left ideal of $RG$ generated by $v$.
Moreover, we say $C(v)$ is a group code of $RG$ with one generator $v$,
since the ring $RG$ is not necessarily a principal ideal ring.
In \cite{Cengellenmis} and \cite{Korban5},
Cengellenmis et al. and Korban et al. constructed reversible codes by utilizing group codes,
and they further constructed some DNA codes with good parameters.
The core construction method is outlined below:

\begin{thm}\cite{Cengellenmis}\label{groupmat}
Let $G$ be a finite group of order $n=2l$ and
$H= \{ e, h_{1}, h_{2}, \ldots , h_{l- 1}\} $ be a subgroup of index $2$ in $G$.
Let $\beta \notin H$ be an element in $G$ with $\beta ^{- 1}=\beta$.
List the elements of $G$ as
\begin{align}
    \{e,h_1,\ldots,h_{l-1},\beta h_{l-1},\ldots,\beta h_1,\beta\},\label{grouporder}
\end{align}
then any linear $G$-code in $R^n$ (a left ideal in $RG$) is a reversible code.
\end{thm}

A matrix construction for composite (matrix) group codes which was first given in \cite{Dougherty7}.
\begin{defn}
Let
$G=\{g_1,g_2,\ldots,g_n\}$ be a group of order $n$,
and let $T_1,T_{2},\ldots,T_{n}$ be finite groups, each of order $k$.
Let $v_{g_i}=\alpha_{(t_i)_1}(t_i)_1+\cdots+\alpha_{(t_{i})_{k}}(t_{i})_{k}\in RT_{i}$
and let $v=\sigma(v_{g_{1}})g_{1}+\cdots+\sigma(v_{g_{n}})g_{n}\in M_{k}(R)G$,
where $\sigma$ is defined as (\ref{gromatr3}).
Then
$$C_k^*(v)=\langle\sigma_k^*(v)\rangle$$
is called composite (matrix) group codes, where
\setlength{\arraycolsep}{1pt}
\renewcommand\arraystretch{0.8}
\[
    \sigma_k^*(v)=
    \begin{pmatrix}
        \sigma(v_{g_1^{-1}g_1})&\sigma(v_{g_1^{-1}g_2})&\ldots&\sigma(v_{g_1^{-1}g_n})\\
        \sigma(v_{g_2^{-1}g_1})&\sigma(v_{g_2^{-1}g_2})&\ldots&\sigma(v_{g_2^{-1}g_n})\\
        \vdots&\vdots&\vdots&\vdots\\
        \sigma(v_{g_n^{-1}g_1})&\sigma(v_{g_n^{-1}g_2})&\ldots&\sigma(v_{g_n^{-1}g_n})
    \end{pmatrix} \in M_n(M_k(R)).
\]
Here the code is generated by taking the all left linear combinations of the rows of the matrix
with coefficients in $R$.
\end{defn}

In \cite{Korban5}, Korban {\it et. al.} constructed reversible codes by utilizing composite group codes,
and further constructed DNA codes with good parameters under different constraints.
The core construction method is outlined below:

\begin{thm}\cite{Korban5}\label{compmatr}
Let $G$ be a finite group of order $n=2l$
and $H=\{e,h_{1},\ldots,h_{l-1}\}$ be a subgroup of index $2$ in $G$.
Additionally, let $\beta \notin H$ be an element in $G$ with $\beta^{-1}=\beta$.
List the elements of $G$ as in (\ref{grouporder}). Next, let $T_{1},T_{2},\ldots,T_{n}$ be finite groups,
each of order $k=2m$ and let $S_{i}=\{e,(s_{i})_{1},(s_{i})_{2},\ldots,(s_{i})_{m-1}\}$ be a subgroup of index $2$ in $T_i$.
Let $\mu_i \notin S_{i}$ be an element in $T_{i}$, with $\mu_i^{-1}=\mu_i$.
List the elements of $T_{i}$ as in the following

\[
    \{e_i,(s_i)_1,\ldots,(s_i)_{m-1},\mu_i(s_i)_{m-1},\mu_i(s_i)_{m-2},\mu_i(s_i)_2,\mu_i(s_i)_1,\mu_i\}.
\]
Then the code $C_k^{*}(v)$ in R$^{kn}$ is a reversible code.
\end{thm}
\begin{remark}
Theorem~\ref{compmatr} is a generalization of Theorem~\ref{groupmat}.
Specifically, Theorem~\ref{compmatr} reduces  to the case of Theorem~\ref{groupmat} when each $T_i$ and $S_i$ are taken as the identity group.
Both methods in Theorem~\ref{compmatr} and Theorem~\ref{groupmat} require the condition that there exists a subgroup of index $2$ in $G$,
and that there exists an element of order $2$ which is not in this subgroup.
However, some groups do not satisfy such conditions.
For example,
a cyclic group of order $n$ (where $4$ divides $n$)
has a subgroup of index $2$ that must contain its unique element of order $2$.
Additionally, when $n \ge 5$, the alternating groups $A_n$ are simple groups
and therefore do not have a subgroup of index $2$,
since a subgroup of index $2$ must be normal.
\end{remark}
We find that for an even-order finite group $G$,
the corresponding group codes can be made reversible by arranging the elements of the group $G$ in an appropriate order.
More generally, all group codes are quasi-cyclic codes.
Specifically, $2$-quasi-cyclic codes can be permuted into reversible codes,
demonstrating the flexibility and potential of this approach.

\begin{thm}
    Let $G$ be a finite group. Then the
group code $C$ of $RG$ is equivalent to a quasi-cyclic code.
More precisely, if the group $G$ contains an element of order $m$,
then the code $C$ is equivalent to a quasi-cyclic code with cyclic length of $m$.

\begin{proof}
    Let $G$ be a finite group with order $ml$ and $g \in G$ with order $m$.
We choose a set of right coset representatives $\{g_1, g_2, ..., g_l\}$ for $G/\langle g \rangle$.
List the elements of $G$ in the following order:
\[
    gg_1, ..., g^mg_1, gg_2,...,g^mg_2,...,gg_l,...,g^mg_l.
\]
Under this ordering of the elements of group $G$,
we define a $R$-linear isomorphism $\Omega: RG \rightarrow R^{ml}$. Then,
for any $(c_{1,1},...,c_{m,1},...,c_{1,n},...,c_{m,n}) \in C$, we have
\[
    v=c_{1,1}gg_1+ \cdots + c_{m,1} g^mg_1 + \cdots + c_{1,l}gg_l + \cdots + c_{m,l}g^mg_l \in \Omega^{-1}(C).
\]
Furthermore,
\[
    gv=c_{m,1}gg_1+ \cdots + c_{m-1,1} g^mg_1 + \cdots + c_{m,l}gg_l + \cdots + c_{m-1,l}g^mg_l \in \Omega^{-1}(C),
\]
and $\Omega(gv) = (c_{m,1},c_{1,1}, ...,c_{m-1,1},...,c_{m,l}, c_{1,n},...,c_{m-1,l}) \in \Omega^{-1}(C)$,
since $\Omega^{-1}(C)$ is a left ideal of $RG$. We are done.
\end{proof}
\end{thm}

\begin{coro}
    Even-order group codes are equivalent to quasi-cyclic codes with cyclic length of $2$.
In particular, even-order group codes are reversible.
\begin{proof}
We just need to order the elements of group $G$ in the following:
\[
    gg_1,gg_2,...,gg_l,g^2g_l,...,g^2g_2, g^2g_1,
\]
where $g \in G$ with order $2$ and $\{g_1, g_2, ..., g_l\}$ is a set of right coset representatives of $G/\langle g \rangle$.
\end{proof}
\end{coro}

The above conclusion seems to imply a confusing inference,
suggesting that cyclic codes are equivalent to quasi-cyclic codes.
However, when discussing cyclic codes,
we actually only focus on their maximum cyclic period while overlooking their shorter cyclic periods.
Next, we will elaborate on this point using a straightforward example to demonstrate that the two concepts are not mutually exclusive.

\setlength{\arraycolsep}{1pt}
\renewcommand\arraystretch{0.8}
\begin{exam}
    The cyclic code with parameters $[6,4,2]$ over $\mathbb{F}_2$,
using $x^2 + x + 1$ as the generator polynomial, has a generator matrix
$\begin{pmatrix}
    1 0 0 0 1 1 \\
    0 1 0 0 1 0 \\
    0 0 1 0 0 1 \\
    0 0 0 1 1 1
\end{pmatrix}$.
    By applying a column permutation
$\begin{pmatrix}
    1 & 2 & 3 & 4 & 5 & 6 \\
    1 & 3 & 5 & 2 & 4 & 6
\end{pmatrix}$,
    we obtain a $2$-quasi-cyclic code with generator $\begin{pmatrix}
        1,1,x \\
        1,1,1 \\
        x,1,1
    \end{pmatrix}$ and generator matrix
$\begin{pmatrix}
    1 0 0 1 0 1 \\
    0 0 1 1 0 0 \\
    0 0 0 0 1 1 \\
    0 1 0 1 0 1 \\
\end{pmatrix}$.
\end{exam}

Thus, we can enhance the two previously mentioned methods for constructing reversible codes to accommodate general even-order groups.
While the proof of this conclusion is omitted here,
we will utilize this approach in Section 5 to construct DNA codes with improved parameters.

\begin{thm} \label{ourConThm1}
    Let $G$ be a finite group of order $n=2l$.
    List the elements of $G$ as
\[
    gg_1,gg_2,...,gg_l,g_l,...,g_2,g_1,
\]
where $g \in G$ with order $2$ and $\{g_1, g_2, ..., g_l\}$ is a set of right coset representatives of $G/\langle g \rangle$.

\noindent 1. Any linear $G$-code in $R^n$ (a left ideal in $RG$) is a reversible code.

\noindent 2. Let $T_{1},T_{2},\ldots,T_{n}$ be finite groups,
each of order $k=2m$ and let $\mu_i \in T_{i}$ with order $2$ for each $i$.
List the elements of $T_{i}$ as in
\[
    \{e_i,(s_i)_1,\ldots,(s_i)_{m-1},\mu_i(s_i)_{m-1},\mu_i(s_i)_{m-2},\mu_i(s_i)_2,\mu_i(s_i)_1,\mu_i\},
\]
where $\{e,(s_{i})_{1},(s_{i})_{2},\ldots,(s_{i})_{m-1}\}$ is a set of right coset representatives of $T_i/\langle \mu_i \rangle$.
Then the code $C_k^{*}(v)$ in $R^{kn}$ is a reversible code.
\end{thm}

It is well known that if an $[n, k, d]$ linear code $C$ over $\mathbb{F}_4$ is reversible,
then its corresponding DNA code satisfies the Hamming distance (HD) constraint and the reverse (R) constraint.
Furthermore, if $C$ also contains the all-one vector,
then its corresponding DNA code satisfies the additional reverse-complement (RC) constraint.

\section{$GC$-weight enumerator of DNA codes}

In DNA storage technology, $GC$-content is an crucial parameter.
$GC$-content is the ratio of guanine ($G$) and cytosine ($C$) bases within a DNA sequence.
Since the chemical bonds between $G$ and $C$ are relatively strong,
the proportion of $G$ and $C$ in a DNA strand impacts its stability and melting temperature.
In DNA storage, it is often desirable to maintain a $GC$-content close to $50\%$ to ensure
high stability of the DNA strand and reduce the likelihood of errors during synthesis and sequencing.
Furthermore, when the $GC$-content of a DNA sequence is $50\%$,
the number of each of the four nucleotides ($A, T, G$, and $C$) in the corresponding DNA molecule is equal,
owing to the double-helical structure of DNA.

In this section, we will establish a connection between the $GC$-weight enumerator of a linear code over $\mathbb{F}_4$ and the Hamming weight enumerator of its trace code,
which will greatly simplify the computation of the number of codewords with a $GC$-content of $n/2$ in a given code of length $n$.
While a preliminary investigation on this issue has been conducted in \cite{Gaborit},
we present here a more direct and explicit description.
Leveraging this result in the subsequent section,
we will derive an efficient algorithm aimed at rapidly obtaining all codewords with $GC$-weight equal to $n/2$ from a given code.
This will facilitate further analysis and applications of such codes.

We consider the field extension $\mathbb{F}_{q^m}/\mathbb{F}_q$, this is a Galois extension of degree
$[\mathbb{F}_{q^m}:\mathbb{F}_q]=m$. Let
\begin{align*}
    \Tr:\mathbb{F}_{q^m} \rightarrow\mathbb{F}_q, \,\,\,  x \mapsto \sum_{i=0}^{m-1} x^{q^i}
\end{align*}
denote the trace mapping. For $\mathbf{a}=(a_1,...,a_n) \in \mathbb{F}_{q^m}^n$, we define
\[
    \Tr(\mathbf{a})= (\Tr(a_1),...,\Tr(a_n)) \in \mathbb{F}_q^n.
\]
In this manner, we obtain an $\mathbb{F}_q$-linear map
\begin{align*}
    \Tr: \mathbb{F}_{q^m}^n  \rightarrow \mathbb{F}_q^n, \,\,\,   \mathbf{a}=(a_1,...,a_n)  \mapsto (\Tr(a_1),...,\Tr(a_n)).
\end{align*}

\begin{defn}
    Let $C$ be an $[n,k]$ linear code over $\mathbb{F}_{q^m}$.

    (a) $C|_{\mathbb{F}_q} =C \cap \mathbb{F}_q^n$ is called the subfield subcode (or the restriction of $C$ to $\mathbb{F}_q$).

    (b) $\Tr(C) = \{\Tr(\mathbf{c})| \mathbf{c} \in C \}$ is called the trace code of $C$.
\end{defn}

The relationship between a code's trace code and its subfield subcode is as follows.
\begin{thm}[Delsarte] \cite{Stichtenoth}
    For a code $C$ over $\mathbb{F}_{q^m}$, we have
\[
    (C|_{\mathbb{F}_q})^{\bot}=\Tr(C^{\bot}).
\]
\end{thm}

\begin{thm} \cite{Stichtenoth}
    Let $C$ be a code of length $n$ over $\mathbb{F}_{q^m}$. Then
\[
    \dim C \le \dim \Tr(C) \le m \cdot \dim C
\]
\end{thm}

We can establish the connection between the $GC$-weight enumerator of a DNA code and the Hamming weight enumerator of its trace code as follows.
\begin{thm}\label{gcrela}
    Let $C$ be a linear code over $\mathbb{F}_4$. Then  we have
\[
    GCW_C(a, b)= \frac{|C|}{|\Tr(C)|} CWE_{\Tr(C)}(a,b).
\]
\begin{proof}
    Consider the map
    \begin{align*}
        \Psi_C : C   \longrightarrow \Tr(C), \,\,\,            (c_1,...,c_n)  \mapsto (\Tr(c_1),...,\Tr(c_n)).
    \end{align*}
    It is easily to verify that this map is a surjective $\mathbb{F}_2$-linear homomorphism.
    Therefore, we have an $\mathbb{F}_2$-linear isomorphism $C/ \ker(\Psi_C) \cong \Tr(C)$.
    Since $\mathbb{F}_4 = \{0, 1, w, w^2\}$, we have $\Tr(0) = 0, \Tr(1) = 0, \Tr(w) = 1, \Tr(w^2) = 1$.
    For any $\mathbf{c} \in C$, we have $n_w(\mathbf{c}) + n_{w^2}(\mathbf{c}) = w(\Psi_C(\mathbf{c}))$,
    thus $w_{gc}(\mathbf{c}) = w(\Psi_C(\mathbf{c}))$.
    Clearly, for any $\mathbf{c} + \ker(\Psi_C) \in C/ \ker(\Psi_C)$,
    every element in the coset $\mathbf{c} + \ker(\Psi_C)$ have the same $GC$-weight.
\end{proof}
\end{thm}

By Delsarte theorem, we can obtain the relationship diagram as follows.
$$
\begin{matrix}
    C & \xrightarrow{trace}& \Tr(C) \\
     \downarrow& & \downarrow\\
     C^{\bot}& \xrightarrow{subfield}& C^{\bot}|_{\mathbb{F}_q}.\\
\end{matrix}
$$
Based on the MacWilliams identity,
we can derive the (complete) weight enumerator of the dual code from the (complete) weight enumerator of a given code.
It is natural to ask if there exists a relationship between the $GC$-weight enumerator of a code and that of its dual code.
The following example shows that there does not exist a general relationship between the $GC$-weight enumerator of a code and that of its dual code.

\setlength{\arraycolsep}{1pt}
\renewcommand\arraystretch{0.8}
\begin{exam}
    Let $C_1$ and $C_2$ be $[18,8,6]$ linear codes over $\mathbb{F}_4=\mathbb{F}_2(w)$, respectively,
where $C_1$ has the generator matrix $[I_8, P_1]$ and $C_2$ has the generator matrix $[I_8, P_2]$.
The matrix $I_8$ is the identity matrix of order 8, and $P_1$ and $P_2$ are given as follows:

    \[ P_1=\begin{pmatrix}
   1  &   1  & w^2  & 1 & w^2 &  w &  w & w^2 &  w &  w \\
   1  & w^2  & 0  & 0  & 1 &  1 &  0 &  0 &  1 &  0 \\
   w^2 &  0  & 1 &  1  & w  & w  & 0 & w^2 &  1 & w^2 \\
   w^2 &  1  & 1 &  w &  w^2  & w &  w &  0 &  1  & 0 \\
   0  &   w & w^2  & w  & w & w^2  & w &  1  & 1 &  w \\
   1 &  w^2 & 0  & 0  & 1 &  1 &  w  & 1 & w^2 &  w \\
   w  & w^2 & w^2 &  w &  w & w^2 &  w  & 1  & 0 &  1 \\
   0 &   1  & 1  & 1  & 0 & w^2 & w^2 & w^2 &  w &  0
    \end{pmatrix},
    P_2 = \begin{pmatrix}
   1 &  0 &   w  & 1  & 1 &  1 &  w  & w  & 0  & w \\
 w^2 & w^2 &  w  & 0 &  w^2 &  1 &  w &  w &  0 &  1 \\
   0 & w^2 & w^2 &  w &  w^2  & 0 & w^2 &  w^2  & w &  0 \\
 w^2 & w^2  &  w & w^2  & 1 &  1  & w  & 0 &  1  & w \\
   0 &  1  & 0  & 1 &  w & w &  w &  w &  0 & w^2 \\
   w &  0 & 0  &  1 &  w & w^2  &  w & w^2 & w^2 &  w \\
   0 &  1 & 0 & w^2 &  1 &  w  &  0  & 1  & 1  & 0 \\
   0 &  1 &  1 &  0 &  1 & w^2  & w  & 0  & 0  & 1
\end{pmatrix}.
\]

By using MAGMA, we can verify that the codes $\Tr(C_1)$ and $\Tr(C_2)$ have the same weight enumerator,
whereas $\Tr(C_1^{\bot})$ and $\Tr(C_2^{\bot})$ have different weight enumerator.
More specifically, $C_1$ and $C_2$ have  same $GC$-weight enumerator,
but $C_1^{\bot}$ and $C_2^{\bot}$ have different $GC$-weight enumerator.
\end{exam}

\section{Algorithms}

In this section, we will leverage previous conclusions to propose a series of algorithms.
These algorithms significantly optimize several key steps in the process of finding DNA codes,
and we demonstrate their effectiveness through several examples of runtime comparisons.
By doing so, we are able to significantly improve the efficiency of finding DNA codes with superior parameters.
For convenience, we define the following notations:
\begin{itemize}
\item{}    $A_{4}^{R,RC}(n,d)$:  the maximum size of DNA codes with Hamming distance, reverse and reverse-complement constraints.
\item{}    $A_{4}^{R,RC,GC}(n,d,\frac{n}{2})$: the maximum size of DNA codes with Hamming distance, reverse, reverse-complement, and $GC$-content constraints.
\item{}    $A_{4}^{R,RC,GC,ffs}(n,d,\frac{n}{2})$: the maximum size of DNA codes with Hamming distance, reverse, reverse-complement, $GC$-content constraints and free from secondary structure.
\end{itemize}
\subsection{Finding of DNA codes}
In \cite{Dougherty6}, many DNA codes with good parameters were constructed  using group codes,
composite matrix group codes.
In Section 3,
we have presented a method to construct reversible codes based on general even-order finite groups.

The classification of finite groups with small orders has already been established.
We can use the function ``SmallGroups($n$)" in the MAGMA software \cite{Bosma} to obtain all the unique (non-isomorphic) groups of a given order
$n$ by entering the order as the input parameter.
For small values of $n$ (specifically, when $n \le 2000$ and $n \ne 1024$), this approach enables us to enumerate all such groups.

Here, we introduce an algorithm designed to efficiently finding DNA codes with optimal or superior parameters.
Our method leverages existing results as input parameters and searches for DNA codes that meet or exceed these criteria.
By incorporating the unique properties of finite groups,
our algorithm aims to identify DNA codes with parameters that are competitive or even superior to those obtained using previous methods.

\begin{algorithm}
    \label{alg1}
    \SetAlgoLined
    \SetKwInOut{Input}{input}\SetKwInOut{Output}{output}
    \Input{Length $n$, distance $d$, $A_4^{RC}(n,d)$, group $G$ with order $n$.}
    \Output{$\mathbf{x}$ and $C_{G,x}$}
    \newcommand{\mycolor}[1]{\color[HTML]{0671b9}{\small #1}}

    \For{$\mathbf{x} \in \mathbb{F}_4^n$ with $wt(\mathbf{x})=d$ }{

        $C_{G,x}$=GroupCode($G$,$x$)\;
        \tcp{\mycolor{Generating a group code $C_{G,x}$ by using Theorem~\ref{ourConThm1}. }}
        \If{$A_4^{RC}(C_{G,x}) > A_4^{RC}(n,d)$ and $d(C_{G,x}) == d$}{
            Save $\mathbf{x}$ and $C_{G,x}$.
        }
    }
    \caption{Find group codes with good parameters}
\end{algorithm}

We will execute the above algorithm on a computer equipped with two AMD EPYC 7542 32-core 2.90 GHz processors.
To expedite the search for optimal parameters,
we will harness the multicore capabilities of the processors by launching multiple threads and assigning distinct parameters to each thread for parallel processing.
However, since the MAGMA programming language lacks an API for multithreading,
we have two viable solutions. One approach involves splitting the different parameters into multiple files for independent execution.
Alternatively, we can leverage a scripting language like Python to invoke MAGMA,
enabling multiprocessing calls and efficient parameter management.

 In the following table, we present some new DNA codes obtained by using Algorithm~\ref{alg1}.
The parameters of these DNA codes are better than current known DNA codes.

\begin{table}[H]
    \fontsize{8pt}{4pt}
    \caption{Lower bounds on $A_{4}^{R,RC}(n,d)$ and $A_{4}^{R,RC,GC}(n,d,\frac{n}{2})$}
        \renewcommand\arraystretch{0.76}
        \begin{tabular}{cccc} \hline
        $n$ & $d$ &  $A_4^{R,RC}(n,d)$ & $\text{Best known}$  \\ \hline
        $32$ & $5$  &  $281474976710656$  & $17592186044416$\cite{Dougherty6}   \\
        $48$ & $5$  &  $4722366482869645213696$  & -  \\
        $48$ & $7$  &  $18446744073709551616$  & -    \\
        $50$ & $3$  &  $309485009821345068724781056$  & $1208925819614629174706176$\cite{Cengellenmis}  \\
        $50$ & $4$  &  $309485009821345068724781056$  &  $302231454903657293676544$\cite{Cengellenmis}   \\
        $64$ & $4$  &  $5192296858534827628530496329220096$  & $20282409603651670423947251286016$\cite{Dougherty6}  \\
        $72$ & $4$  &  $21267647932558653966460912964485513216$  & $5192296858534827628530496329220096$\cite{Cengellenmis} \\
        $96$ & $4$  &  $374144419156711147060143317175368453031918731001856$  & - \\
    \hline
    $n$ & $d$ &$A_4^{R,RC,GC}(n,d,\frac{n}{2})$ & $\text{Best known}$  \\ \hline
        $32$ & $5$  &  $78784808878080$ & $4924788572160$\cite{Dougherty6} \\
        $48$ & $5$  &  $541026086295649124352$  & - \\
        $48$ & $7$  &  $4226766299184758784$ & - \\
        $50$ & $3$  &  $34747482913131001584549888$  & - \\
        $50$ & $4$  &  $34747482913131001584549888$  & - \\
        $64$ & $4$  &  $1031675674782403258882349605060608$ & $4029983108584131998207975620608$\cite{Dougherty6} \\
        $72$ & $4$  &  $3985801709120774881778580468417953792$ & -\\
        $96$ & $4$  &  $60777489562618370865292269687095803826096339681280$ & - \\ \hline
    \end{tabular}
\end{table}

\subsection{$50\%$ $GC$-content constraint}

To maintain stability in DNA molecular chains utilized for information storage,
it is essential to maintain a $GC$-content of $50\%$ for each DNA molecular chain.
Therefore, we need to calculate the number of codewords from a code that satisfy the $50\%$ $GC$-content constraint.
In Section~4, we have established a relationship between the $GC$-weight enumerator of a
code over $\mathbb{F}_4$ and the Hamming weight enumerator of its trace code,
which enables a more efficient computation process.
Due to the large dimensions of reversible codes involved in the current DNA code constructions,
the dimensions of their dual codes are relatively small. To further optimize the computation,
we approach the problem from the perspective of the weight enumerator of the dual code.
By utilizing the MacWilliams relationship and the specific properties of Krawtchouk polynomials,
we can more efficiently calculate the number of codewords  with $50\%$ $GC$-content from a code.
\begin{thm}\cite{MacWilliams}
    For any binary $[n,k]$ linear code $C$, we have
\[
    A_s = \frac{1}{|C^{\bot}|}\sum_{i=0}^nA_i'P_s(i),
\]
where  $\{A_i\}$ and $\{A_i'\}$ $(i=0,...,n)$ denote the weight spectrum of the
code $C$ and its dual code, respectively, and the Krawtchouk polynomial $P_s(x;n)=$
    $P_s(x)$ is defined by
    $$P_s(x;n)=\sum_{j=0}^s(-1)^j\binom{x}{j}\binom{n-x}{s-j},\quad s=0,1,2,\ldots $$
\end{thm}

The following lemma is easily to prove, for the sake of integrity, we provide a proof here.
\begin{lem}
Let $n = 2s$ and $x$ be odd. Then $P_s(x;n)=0$.

\begin{proof}
    $$P_s(x;2s)=\sum_{j=0}^s(-1)^j\binom{x}{j}\binom{2s-x}{s-j}.$$
When $x \le s$,
\begin{align*}
    (-1)^j\binom{x}{j}\binom{2s-x}{s-j} &= 0, \ \text{for} \ x< j \le s,  \\
    (-1)^j\binom{x}{j}\binom{2s-x}{s-j} + (-1)^{x-j}\binom{x}{x-j}\binom{2s-x}{s-(x-j)}&=0, \ \text{for} \ 0 \le j \le x.
\end{align*}
When $x > s$,
\begin{align*}
    (-1)^j\binom{x}{j}\binom{2s-x}{s-j} &= 0, \ \text{for} \ 0 \le j < x-s, \\
    (-1)^j\binom{x}{j}\binom{2s-x}{s-j} + (-1)^{x-j}\binom{x}{x-j}\binom{2s-x}{s-(x-j)}&=0, \ \text{for} \ x-s \le j \le s.
\end{align*}
Thus $P_s(x;2s) = 0$.
\end{proof}
\end{lem}

Therefore, we only need to calculate the weight spectrum of the even-weight subcode of code $C^{\bot}$.
We have the following results:

\begin{coro}\label{Krawpolres}
    For any binary $[n,k]$ linear code $C$, we have
\[
    A_{\frac{n}{2}} = \frac{1}{|C^{\bot}|}\sum_{i=0}^{n/2}A_{2i}''P_{\frac{n}{2}}(2i),
\]
where $\{A_j''\}$ $(j=0,...,n)$, denote the weight
spectrum of the even weight subcode of $C^{\bot}$.
\end{coro}

Combining the conclusion of Theorem~\ref{gcrela},
we can easily arrive at the following conclusion.
\begin{coro}
    For any $[n,k]$ linear code $C$ over $\mathbb{F}_4$, the number of codewords $N_{\frac{n}{2}}(C)$ with $GC$-weight equal to $n/2$ is given by
\[
    N_{\frac{n}{2}}(C) = \frac{|C|}{2^n}\sum_{i=0}^{n/2}A_{2i}'''P_{\frac{n}{2}}(2i),
\]
where $\{A_j'''\}$ for $j = 0, 1, \ldots, n$, denote the weight spectrum of the even weight subcode of the dual code $\Tr(C)^{\bot}$.
\end{coro}

During the actual computation process,
we employ an optimization strategy where we precompute the values of $P_s(2i)$ and directly retrieve these precomputed values whenever needed,
instead of recalculating them repeatedly.
This approach significantly reduces the performance overhead caused by redundant calculations of $P_s(2i)$.
For the method of calculating $P_s(2i)$, please refer to reference \cite{OurData}, which contains the specific values of $P_s(2i)$ and the calculation program for the Krawtchouk polynomial written by MAGMA.

\begin{table}[H]
    \caption{Performance comparison}
    \begin{center}
    {  \footnotesize
        \renewcommand\arraystretch{1.2}
        \begin{tabular}{ccccc|ccccc} \hline
        $n$ & $k$ &  $t_0$ & $t_1$ & $t_2$  & $n$ & $k$ &  $t_0$ & $t_1$ & $t_2$ \\ \hline \hline
        30 & 11   & 17.660  & 0.047 & 0.016 & 70 & 22  & - & 34.969 & 17.313 \\
        40 & 10   & 5.940  & 0.375 & 0.203 & 70 & 24  & - & 2.313 & 1.187 \\
        40 & 16   & 23357.800  & 0.078 &  0.031 & 70 & 26  & - & 0.265 & 0.188 \\
        50 & 14   &  1529.380 & 1.094 & 0.59 & 70 & 28  & - & 0.203 & 0.125 \\
        50 & 20   & - &  0.094 & 0.062 & 70 & 30  & - & 0.204 & 0.125 \\
        60 & 16   &  29781.250 &  78.625 & 67.938  & 80 & 24  & - & 6120.078 & 1129.234 \\
        60 & 18   &  - & 8.688 & 4.375 & 80 & 26  & - & 143.484 & 71.313 \\
        60 & 20   & -  & 0.562 & 0.329 & 80 & 28  & - & 9.907 & 5.000 \\
    \hline
    \end{tabular}}
    \end{center}
    \begin{threeparttable}
        \begin{tablenotes}
            \footnotesize
            \item[1]
            In this table, `$n$' represents the code length, `$k$' represents the code dimension,
            `$t_0$' represents the time consumed for computing the complete weight enumerator,
            `$t_1$' represents the time consumed for calculating the $GC$-weight enumerator using Theorem~\ref{gcrela},
            and `$t_2$' represents the time consumed for computation using Theorem~\ref{Krawpolres}.
        \end{tablenotes}
    \end{threeparttable}
\end{table}

\begin{remark}
In Table 2, we compare the time consumption of three different computational methods for random codes with length $n$ and dimension $k$ over $\mathbb{F}_4$, repeated $100$ times.
Through comparative analysis, we draw the following conclusions:

1. The method used in Theorem~\ref{gcrela} demonstrates a significant efficiency improvement compared to directly computing the complete weight enumerator.
This improvement holds true for any given parameters $n$ and $k$,
indicating that Theorem~\ref{gcrela} provides an efficient solution.

2. Compared to Theorem~\ref{gcrela}, the method utilizing Theorem~\ref{Krawpolres} does exhibit higher execution efficiency under certain specific parameter settings.
However, this efficiency gain is not universal,
as its effectiveness may not be apparent for other parameter combinations.
Here, we present only a few positive examples to illustrate this point.

3. For some parameter combinations, especially when $n$ and $k$ are large,
directly computing the complete weight enumerator can become extremely time-consuming,
and in some cases, even infeasible (denoted by `-' in the table).
This further underscores the importance of using efficient algorithms such as Theorem~\ref{gcrela} and Theorem~\ref{Krawpolres}.
\end{remark}

\subsection{Other specific constraints}

Previously, we discussed the construction of codes that satisfy the reverse-complement distance and the complement distance,
as well as the calculation of the number of codewords in a code with a $GC$-content of $n/2$.
Here, we present a specific algorithm to obtain all codewords with a $GC$-content of $n/2$ from a given code.
Additionally, we provide two algorithms: one to identify whether a DNA string is free from secondary structures,
and another to assess whether it is conflict-free.

Let $C$ be an $[n,k]$ linear code over $\mathbb{F}_4=\mathbb{F}_2(w)$ with a generator matrix
$G = [\mathbf{g}_1,...,\mathbf{g}_k]^T$.
If we consider the code $C$ as a vector space over $\mathbb{F}_2$,
then the set $\{\mathbf{g}_1,...,\mathbf{g}_k, w\mathbf{g}_1,...,w\mathbf{g}_k\}$ is linearly independent over $\mathbb{F}_2$ and forms a basis for the vector space $C$ over $\mathbb{F}_2$.

We have a surjective $\mathbb{F}_2$-linear homomorphism
\begin{align*}
    \Psi_C: C  \longrightarrow \Tr(C), \,\,\,    \mathbf{c}  \mapsto \Tr(\mathbf{c}),
\end{align*}
and $\ker(\Psi_C) = C|_{{\mathbb{F}_2}}$.

Therefore, the vectors $\Tr(\mathbf{g}_1),...,\Tr(\mathbf{g}_k),\Tr(w\mathbf{g}_1),...,\Tr(w\mathbf{g}_k)$ generate $\Tr(C)$.
From these, we can find a basis denoted as $\Tr(\mathbf{t}_1),...,\Tr(\mathbf{t}_{k_1})$ for $\Tr(C)$ where $\{\mathbf{t}_1,...,\mathbf{t}_{k_1}\}$ is a subset of $\{\mathbf{g}_1,...,\mathbf{g}_k,w\mathbf{g}_1,...,w\mathbf{g}_k\}$ and $k_1$ is the dimension of $\Tr(C)$.
Furthermore, we have an $\mathbb{F}_2$-linear isomorphism
\begin{align*}
   \hat{\Psi}_C: \langle \mathbf{t}_1,...,\mathbf{t}_{k_1} \rangle  \longrightarrow \Tr(C), \,\,\,   \mathbf{c}  \mapsto \Tr(\mathbf{c}).
\end{align*}

We denote the set of codewords in $\Tr(C)$ with weight $n/2$ as $W_{\frac{1}{2}}(C)$.
Then, the set of codewords in $C$ with $GC$-weight $n/2$ is given by $\hat{\Psi}_C^{-1}(W_{\frac{1}{2}}(C)) + \ker(\Psi_C)$.

\begin{algorithm}
    \label{algobb}
    \SetAlgoLined
    \SetKwInOut{Input}{Input}\SetKwInOut{Output}{Output}
    \Input{Linear code $C$ over $\mathbb{F}_4$.}
    \Output{$S_2$ and $C|\mathbb{F}_2$}
    \newcommand{\mycolor}[1]{\color[HTML]{0671b9}{\small #1}}
    $G$ := GeneratorMatrix($C$)\;
    \tcp{\mycolor{Obtain a generator matrix $G=[\mathbf{g}_1,...,\mathbf{g}_k]^T$ of $C$.}}
    $k$ = Dimension($C$)\;
    $k_1$ = Dimension(Tr($C$))\;
    $T$ = []\;
    $T_1$ = []\;
    \For{$i=1,...,2k$ }{
        \eIf{i <= n}{
            Append($T$, $\mathbf{g}_i$)\;
            Append($T_1$, Tr($\mathbf{g}_i$))\;
            \If{CheckLinearIndependentOverF2($T_1$) == false}{
                Pop($T$, $\mathbf{g}_i$)\;
                Pop($T_1$, Tr($\mathbf{g}_i$))\;
            }
        }{
            Append($T$, $w\mathbf{g}_i$)\;
            Append($T_1$, Tr($wg_i$))\;
            \If{CheckLinearIndependentOverF2($T_1$) == false}{
                Pop($T$, $w\mathbf{g}_i$)\;
                Pop($T_1$, Tr($w\mathbf{g}_i)$)\;
            }
        }
        \If{Size($T$) ==  $k_1$}{
            break\;
        }
    }
    $S$ = Words(Tr($C$), $\frac{n}{2}$)\;
    \tcp{\mycolor{Obtain all codewords of $\Tr(C)$ with weight equal to $n/2$.}}
    $S_2$ = []\;
    \For{$\mathbf{c} \in S$ }{
        $\mathbf{x}$ = Sovle($T_1, \mathbf{c}$)\; \tcp{ \mycolor{Obtain a verctor $\mathbf{x}$ satisfy $\mathbf{x}*T_1 = c$.}}
        Append($S_2, \mathbf{x}*T$)\;
    }
    \Return{$S_2$, SubfieldSubcode($C$)}
    \caption{Obtain all codewords of $C$ with $GC$-weight equal to $n/2$.}
\end{algorithm}

Based on the above analysis, we provide Algorithm~\ref{algobb} based on this approach to help us obtain the set of codewords in code $C$ with a $GC$-weight of $n/2$ more efficiently.

\begin{table}[H]
    \caption{Running time comparison}
    \begin{center}
    {  \footnotesize
        \renewcommand\arraystretch{1.2}
    \begin{tabular}{ccccc|ccccc} \hline
        $n$ & $d$ &  $A_4^{R,RC,GC}(n,d,\frac{n}{2})$ & $t_0$ & $t_1$ & $n$ & $d$ &  $A_4^{R,RC,GC}(n,d,\frac{n}{2})$ & $t_0$ & $t_1$ \\ \hline \hline
        10 & 4  &  $1008$  & $0.015$  & $0.014$ & 16 & 8  &  $4800$  & $0.094$  &  $0.016$  \\
        12 & 4  &  $29568$  & $0.265$  & $0.016$ & 18 & 4  &  $3153920$  &  $92.156$ & $0.110$ \\
        12 & 6  &  $1848$  &  $0.031$ & $0.016$ & 18 & 6  &  $204800$  &  $6.406$ & $0.031$ \\
        16 & 4  &  $125952$  &  $1.313$ & $0.016$ & 20 & 4  &  $378380288$  &  $92.859$ &  $2.188$  \\
    \hline
    \end{tabular}}
    \end{center}
    \begin{threeparttable}
        \begin{tablenotes}
            \footnotesize
            \item[2]
            In this table,
            `$t_0$' represents the time consumed for iterating through the codewords of a given code,
            `$t_1$' represents the time consumed for algorithm~\ref{algobb}.
        \end{tablenotes}
    \end{threeparttable}
\end{table}
We present the comparison results between the running time of our algorithm~\ref{algobb}
and the running time of iterating through the codewords of a code to obtain the codewords with $n/2$ $GC$-weight.
The comparison results can be seen in Table~3.

The research in \cite{Bornholt}, \cite{Erlich}, \cite{Song} indicates that good DNA codes should exclude consecutive nucleotide repetitions like ($GC\mathbf{TTTT}AC$) due to insertion/deletion errors.
\cite{Jain} and \cite{Kova} suggest avoiding blocks like $\mathbf{GCTGCT}AG$.
\cite{Benerjee} explores conflict-free DNA strings devoid of homopolymers and consecutive nucleotide block repeats.
\cite{Kari},\cite{Milenkovic}, \cite{Nelms} advise against pairs of substrings whose reverse complements form each other,
as this can lead to anti-parallel double stranded hairpin structures.

A DNA string is deemed free from secondary structures (stem length $> 2$) if it lacks any pair of substrings $>2$ bases long,
where one is the reverse-complement of the other. This ensures no secondary structures form,
preserving the chemical activity crucial for DNA computing.
Examples of such considerations are found in \cite{Bornholt}, \cite{Benerjee}, \cite{Kari}, and \cite{Kova},
which also emphasize avoiding consecutive repetitions of identical substrings.

\begin{algorithm}
    \SetAlgoLined
    \SetKwInOut{Input}{Input}\SetKwInOut{Output}{Output}
    \Input{DNA string $D_s$}
    \Output{Bool value}
    $l =$ Length($D_s$)$/2$\;
    \For{i=1...l}{
        s = l-i+1\;
        t = n-2*s\;
        \For{j=0...t}{
            \If{$D_s[j+1:j+s] == D_s[j+s+1:j+2*s]$}{
                \Return{false\;}
            }
        }
    }
    \Return{true\;}

    \caption{Determination of $n/2$ conflict-free DNA codeword.}
\end{algorithm}

\begin{algorithm}
    \label{alg4}
    \SetAlgoLined
    \SetKwInOut{Input}{Input}\SetKwInOut{Output}{Output}
    \Input{DNA string $D_s$}
    \Output{Bool value}
    n = Length($D_s$)\;
    \For{i=1...(n-2)}{
        $dss=D_s[i:i+2]$\;
        rc=ReverseComplement(dss)\;
        \For{j=i...(n-2)}{
            $rss=D_s[j:j+2]$\;
            \If{rc == rss}{
                \Return{false\;}
            }
        }
    }
    \Return{true\;}

    \caption{Determination of DNA codeword  free from secondary structure.}
\end{algorithm}
At the end of this paper,
we present two determination algorithms specifically designed to ascertain whether a DNA string is
free from secondary structures or is conflict-free.
Leveraging Algorithms \ref{alg1},\ref{algobb} and~\ref{alg4},
we have efficiently identified some DNA codes with good parameters that are devoid of secondary structures.
We present these codes in Table 4.
All the data and most of the algorithms presented in this paper can be found in \cite{OurData}.

\begin{table}[H]
    \caption{Lower bounds on $A_{4}^{R,RC,GC}(n,d,\frac{n}{2})$ and $A_{4}^{R,RC,GC,ffs}(n,d,\frac{n}{2})$}
    \begin{center}
    {  \footnotesize
        \renewcommand\arraystretch{1.2}
        \begin{tabular}{ccccccc} \hline
        $n$ & $d$ & group &  $A_4^{R,RC,GC,ffs}(n,d,\frac{n}{2})$ & Best known & $A_4^{R,RC,GC}(n,d,\frac{n}{2})$ & Best known  \\ \hline \hline
        10 & 4  & $1$ &  $676$  & $-$  & $1008$ &  $-$ \\
        12 & 4  & $3$ &  $14316$  & $-$  & $29568$ &  $14784$\cite{Aboluion1} \\
        12 & 6  & $3$ &  $796$  &  $-$ & $1848$  & $1848$\cite{Cengellenmis} \\
        16 & 4  & $2$ &  $29064$  &  $-$ &  $125952$ & $-$ \\
        16 & 6  & $4$ &  $7308$  &  $-$ &  $26720$ & $-$ \\
        16 & 8  & $4$ &  $1384$  &  $-$ & $4800$  & $-$ \\
        18 & 4  & $3$ &  $554760$  &  $-$ & $3153920$ & $-$ \\
        18 & 6  & $3$ &  $38320$  &  $-$ &  $204800$ & $-$ \\
        20 & 4  & $3$ &  $43092124$  &  $-$ &  $378380288$ & $-$ \\
        20 & 6  & $1$ &  $173864$  &  $-$ &  $1478048$ & $-$ \\
    \hline
    \end{tabular}}
    \end{center}
    \begin{threeparttable}
        \begin{tablenotes}
            \footnotesize
            \item[3]
            In this table, the value $i$ of "group" represents the group of order $n$ obtained by using the Magma function SmallGroup($n$, $i$).
        \end{tablenotes}
    \end{threeparttable}
\end{table}

\section{Conclusion and further research}

In this paper, firstly, we prove that group codes constitute a special class of quasi-cyclic codes.
As a result, we extend the results of constructing reversible group codes \cite{Cengellenmis} and reversible composite group codes \cite{Korban5} to general even-order finite groups.
Using these improved results,
we present parallel searching algorithms to discover novel DNA codes with superior parameters.
Secondly,
we establish a relationship between the $GC$-weight enumerator of DNA codes and the Hamming weight enumerator of their trace codes,
which greatly improves the computational efficiency of searching for DNA codes.
Furthermore, this relationship provides a more practical and intuitive approach compared to previous methods \cite{Gaborit},
particularly for generating DNA codes with $50\%$ $GC$-content.
Leveraging this relationship, we propose an efficient algorithm for generating DNA codes with $50\%$ $GC$-content.
Finally, we present algorithms to determine whether a DNA code is free from secondary structures or conflict-free,
and introduce some new DNA codes with improved parameters under multiple constraints.

In \cite{Barg}, A. M. Barg and I. I. Dumer proposed a more efficient algorithm specifically designed for rapidly calculating the Hamming  weight enumerator
of a given weight within binary cyclic codes.
Since cyclic codes are a special subclass of group codes,
it is a worthwhile question to explore whether similar optimization algorithms exist for binary group codes.
In the context of constructing DNA codes with a $GC$-content of $50\%$,
such optimization algorithms are particularly crucial.

\vskip 3mm
\noindent{\bf Acknowledgement.}

This work was supported by National Natural Science Foundation of China under Grant Nos.12271199 and 12171191 and The Fundamental Research Funds for the Central Universities 30106220482.


\end{document}